\documentclass[11pt,prl,showkeys,onecolumn,showpacs]{revtex4}
\usepackage{slashbox}
\usepackage{amsmath,graphicx,amsfonts,bm,amssymb}
\usepackage{times}
\begin{document}
%\begin{CJK*}{GBK}{song}
\title{Fast geometric gate operation of superconducting charge qubits in circuit QED}

\author{Zheng-Yuan Xue}

\affiliation{Laboratory of Quantum Information Technology, and
School of Physics and Telecommunication Engineering, South China
Normal University, Guangzhou 510006,  China}

\date{\today}

\begin{abstract}
A scheme for coupling superconducting charge qubits via a
one-dimensional superconducting transmission line resonator is
proposed. The qubits are working at their optimal points, where they
are immune to the charge noise and  possess long decoherence time.
Analysis on the dynamical time evolution of the interaction is
presented, which is shown to be insensitive to the initial state of
the resonator field. This scheme enables fast gate operation and is
readily scalable to multiqubit scenario.
\end{abstract}



\maketitle

Quantum computers have been paid much attention in the past decade
and solid state systems are promising candidates for novel scalable
quantum information processing \cite{youpt}. In particularly, the
idea of placing superconducting qubits inside a  cavity, i.e., the
circuit quantum electrodynamics (QED), has been illustrated
\cite{yale,yaledetune} to have several practical advantages
including strong coupling strength, immunity to noises, and
suppression of spontaneous emission.

Decoherence always occur during real quantum evolutions and
therefore stands in the way of physical implementation of quantum
computers. So, how to suppress this infamous decoherence effects is
a main task for scalable quantum computation. To fight against
cavity decay, in typical circuit \cite{yaledetune} and  cavity
\cite{zhengdetune} QED systems, convectional wisdom is to resort to
the so-called large detuning interaction method. Similarly, in
trapped thermal ions system, the famous bichromatic excitation
scheme \cite{sm1} couples ions by virtue excitation of phonon mode,
also uses the large detuning interaction. Later investigation shows
that the logical operation obtained is of the geometric nature
\cite{sm2} and therefore has high fidelity \cite{f}. Meanwhile, it
is shown that by periodically decoupling to the common phonon  mode,
the large detuning constrain can be removed \cite{sm2} so that fast
gate operation can be achieved \cite{zhengdecoup}. Similar strategy
can be adopted in cavity QED system with strong driven atoms
\cite{zhengcavity}, superconducting charge qubits in a microwave
cavity by introducing ac magnetic flux \cite{zhu} and
superconducting flux qubits inductively coupled to a common
resonator \cite{wangyd}.

In typical circuit QED system, up to now, theory and experimental
explorations are still in the stage of large detuning interaction.
Here, we propose to coupe superconducting charge qubits via a
one-dimensional (1D) superconducting transmission line resonator
(cavity). The qubits are capacitively coupled to the 1D
superconducting cavity \cite{yale} and work at their optimal points,
where they are immune to the charge noise and possess long
decoherence time. The gate operation is shown to be insensitive to
the initial state of the cavity field, and thus greatly suppress the
decoherence effect  from the cavity decay. This scheme removes the
requirement of large detuneing, and thus enables fast gate
operation. Finally, the solid-state set-up is readily scalable to
multiqubit scenario.

Before proceeding, we would like to explain our proposal in a more
physical way. Usually, 2-qubit coupling is demonstrated with large
qubit-cavity detuning $\delta\gg g$, e.g., in Ref. \cite{yale},
which makes the coupling quite weak. In this regime, there is no
energy exchange between qubits and cavity. The effective coupling of
energy conservation transitions can be determined by second-order
perturbation theory \cite{sm2}. Meanwhile, the coupling usually
contains cavity-state-dependent energy shift, i.e.,
$a^{\dagger}a\sigma^z_j$ terms. Outside this regime the internal
state is strongly entangled with the cavity state during the gate
operation. By adding a driven field to the cavity field, we get
effective driven for all qubits, which is similar to scheme of
atomic cavity QED with strong driven \cite{s}. This driven field
further  mix both the cavity and the qubit state. For successful
gate operation (evolution independent of the cavity state) we have
to ensure that the cavity returns to its initial state at the end of
the gate. Fortunately, this  can be achieved  by appropriately
chosen parameters, i.e., to fulfill Eq. (\ref{parameter}), where the
two mix mechanisms of the cavity field state conceal each other and
thus result in cavity state independent evolution of the system.

\begin{figure}[bp]
\includegraphics[width=9cm]{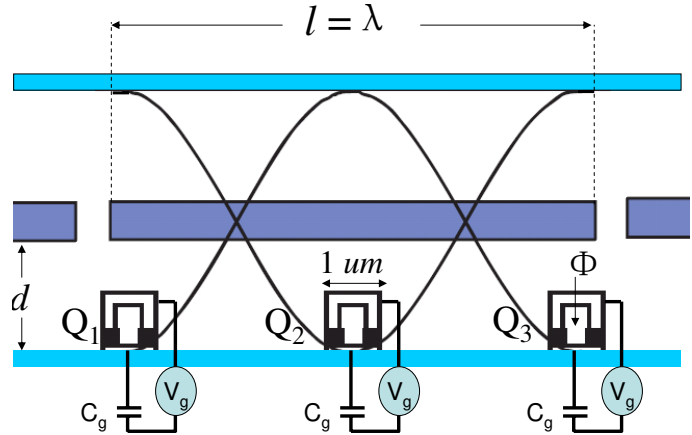}
\caption{(Color online) The architecture  of 3 superconducting
charge qubits capacitively coupled to a 1D cavity, which consists of
a full-wave section ($l=\lambda$) of superconducting coplanar
waveguide. Qubits are placed between the superconducting lines and
is capacitively coupled to the center trace at a maximum of the
voltage standing wave (solid cosine curves), and thus yielding
maximum coupling. Qubit consists of two small  Josephson junctions
in a 1 $\mu$m loop to permit tuning of the effective Josephson
energy by an external flux $\Phi$. Input and output signals can be
coupled to the resonator, not shown here,via the capacitive gaps in
the center line.} \label{setup}
\end{figure}

Fig. (\ref{setup}) shows our proposed setup with 3 qubits. The 1D
cavity consists of a full-wave section ($l=\lambda$) of
superconducting coplanar waveguide. A distinct advantage of circuit
QED approach is that the zero-point energy is distributed over a
very small effective volume, which leads to significant rms voltages
between the center conductor and the adjacent ground plane at the
antinodal positions. At a cavity resonant frequency of 10 GHz and
$d=10$ $\mu$m, $V_{rms}=2$ $\mu$V corresponding to electric fields
$E_{rms}=0.2$ V/m, which is several hundreds times larger
\cite{yale} than that of achieved in 3D cavity with Rydberg atoms.

The superconducting charge qubit considered here consists of a small
superconducting box with excess Cooper-pair charges, formed by an
symmetric SQUID with the capacitance $C_{J}$ and Josephson coupling
energy $E_J$, pierced by an external magnetic flux $\Phi$, permit
tuning of the effective Josephson energy. A control gate voltage
$V_g$ is connected to the system via a gate capacitor $C_g$.  Focus
on the charge regime ($E_J\ll E_c=2e^2/C_\Sigma$ with
$C_\Sigma=C_g+2C_{J}$), at temperatures much lower than the charging
energy ($E_{c}$) and restricting the induced charge [$\bar{n}=C_g
V_g/(2e)$] to the range of $\bar{n}\in [0,1]$, only a pair of
adjacent charge states $\{|0\rangle,|1\rangle\}$ on the island are
relevant. Then, the device is described by \cite{youpt}
\begin{equation}
\label{h1r} H_s=-{\epsilon \over 2} \widetilde{\sigma}^z-{\Delta
\over 2} \widetilde{\sigma}^x,
\end{equation}
where $\epsilon =E_c(1-2\bar{n})$, $\Delta=2E_J\cos(\pi\Phi/\phi_0)$
with $\phi_0$ being the flux quanta, $\widetilde{\sigma}^x$ and
$\widetilde{\sigma}^z$ are the Pauli matrices in the
$\{|0\rangle,|1\rangle\}$ basis. Note that the qubit splitting now
can be tunable by the external magnetic flux, which can be used to
turn on/off  cavity mediated qubits interaction. As this can be
tuned individually, the cavity mediated qubits interaction can be
implemented on selective qubits.

The qubits are capacitively coupled to the  cavity, as shown in Fig.
(\ref{setup}). For simplicity, we here assume that the cavity has
only a single mode that plays a role. To obtain maximum coupling
strength, they are fabricated close to the voltage antinodes of the
cavity. As the wave length of the cavity mode ($\lambda\sim 1$ cm)
is much larger than the linear length of the qubit (1 $\mu$m), we
may treat the qubit-cavity coupling  as a constant within the qubit
geometry. This coupling is determined by the gate voltage, which
contains both the dc contribution and a quantum part. If the qubit
is placed in the center of the resonator, as shown in Fig.
(\ref{setup}), the latter part contribution is given by
$V_{rms}^0(a+a^\dagger)$ with $V_{rms}^0$ being the rms voltage
between the center conductor and the ground plane. Then, the
Hamiltonian describes this setup takes the form of \cite{yale}
\begin{equation}
\label{h1c} H_c=\omega_r a^\dag a- \sum_{j=1}^{N}\left[{\epsilon
\over 2} \widetilde{\sigma}^z_j+{\Delta \over 2}
\widetilde{\sigma}^x_j+g_j(a+a^\dagger)(1-2\bar{n}-\widetilde{\sigma}^z_j)\right],
\end{equation}
where we have assume $\hbar=1$, $\omega_r$, $a$ and $a^\dag$ is the
frequency, annihilation and creation operator of the cavity field,
respectively; the coupling strength of $j$th qubit to the cavity is
$g_j=eC_{g,j}V_{rms}^0/ C_{\Sigma,j}\in[5.8, 100]$ MHz \cite{yale}.
Rotate to the qubit eigen basis of
\begin{eqnarray}
|\uparrow\rangle=\cos{\theta \over 2}|0\rangle+\sin{\theta \over
2}|1\rangle,\quad |\downarrow\rangle=-\sin{\theta \over
2}|0\rangle+\cos{\theta \over 2}|1\rangle,
\end{eqnarray}
where  $\tan\theta=\Delta/\epsilon$,  Hamiltonian (\ref{h1c}) takes
the form of
\begin{equation}
\label{h1c2} H_c=\omega_r a^\dag a+\sum_{j=1}^{N}\left[{\omega_a
\over 2}
\sigma^z_j-g_j(a+a^\dagger)(1-2\bar{n}-\cos\theta\sigma^z_j+\sin\theta\sigma^x_j)\right],
\end{equation}
where $\omega_a=\sqrt{\epsilon^2+\Delta^2}$ is the qubit splitting,
$\sigma^x$ and $\sigma^z$ are the Pauli matrices in the
$\{|\uparrow\rangle,|\downarrow\rangle\}$ basis. The qubits are set
to work at their optimal points ($\bar{n}=1/2$, $\omega_a=\Delta$
and $\theta=\pi/2$, corresponding to $\widetilde{\sigma}_x$ basis),
where they are immune to the charge noise and possess long
decoherence time. Then, the Hamiltonian (\ref{h1c2}) reduces to
\begin{equation}
\label{h1c3} H_c=\omega_r a^\dag a+ \sum_{j=1}^{N}\left[{\Delta
\over 2} \sigma^z_j-g_j(a+a^\dagger) \sigma^x_j\right].
\end{equation}
Neglecting fast oscillating terms using the rotating-wave
approximation lead Hamiltonian (\ref{h1c3}) to the usual
Jaynes-Cummings form
\begin{equation} H_{JC} = \omega_r a^\dag a +
\sum_{j=1}^{N}\left[{\Delta_{j}  \over 2}\sigma^z_{j} -  g_j
\left(a^\dag \sigma^-_{j}+a\sigma^+_{j} \right)\right].
\end{equation}

Meanwhile, driving in the form of
\begin{equation}\label{hd}
h=\varepsilon(t)a^\dagger e^{-i\omega_d t}+ \varepsilon^*(t) a
e^{i\omega_d t}
\end{equation}
on the resonator can be obtained \cite{yale} by capacitively
coupling it to a microwave source with frequency $\omega_d$ and
amplitude $\varepsilon(t)$. Depending on the frequency, phase, and
amplitude of the drive, different logical operations for qubit can
be realized. To get fast gate, we work with large amplitude driving
fields, where quantum fluctuations are very small compare with the
drive amplitude, and thus the drive can be considered as a classical
field. In this case, it is convenient to displace the field
operators using the time-dependent displacement operator
\cite{scully}:
\begin{equation}\label{dis}
D(\alpha) = \exp\left(\alpha a^\dagger - \alpha^* a \right).
\end{equation}
Under this transformation, the field $a$ goes to $a+\alpha$ where
$\alpha$ is a c-number representing the classical part of the field.
Choosing
\begin{equation}
i\dot\alpha = \omega_r\alpha +\varepsilon(t) \exp({-i\omega_d t})
\end{equation}
to eliminate the direct drive on the resonator, then
the displaced Hamiltonian reads \cite{yale}
\begin{equation}\label{hd}
 H_D=\omega_r a^\dagger a +
\sum_{j=1}^{N}\left\{{\Delta_{j}  \over 2}\sigma^z_{j} -  g_j\left[
\left(a+\alpha\right)\sigma^+_{j} + \text{H.c}.\right]\right\}.
\end{equation}
Under resonant driving ($\Delta_j=\omega_d$), the drive amplitude is
independent of time, and change to a frame rotating at $\omega_d$,
the displaced Hamiltonian reads
\begin{equation}\label{rotate}
H_\mathrm{RF} = \delta a^\dag a + \sum_{j=1}^N \left[{\Omega\over 2}
\sigma_x^j- g_j \left(a \sigma^{+}_j +a^\dag \sigma^{-}_j
\right)\right].
\end{equation}
where $\delta=\omega_r-\omega_d$ and $\Omega=2g\varepsilon/\delta$
is the Rabi frequency.

Rotate to the eigen basis of $\sigma_x$
\begin{eqnarray}
&& H_{RF} = H_0+H_{int}, \notag\\
&& H_0=\delta a^\dag a +{\Omega\over 2} \sum_{j=1}^N \sigma_z^j, \\
&& H_{int}= - {1 \over 2} \sum_{j=1}^N \left[g_j a
\left(\sigma_{z}^j +|+\rangle_j\langle-|-|-\rangle_j\langle+|\right)
+\text{H.c.}\right]\notag.
\end{eqnarray}
In the interaction picture with respect to $H_0$ the interaction
Hamiltonian reads \cite{s}
 \begin{eqnarray} H_{1}=- {1 \over 2} \sum_{j=1}^N g_j a  e^{-i\delta t}\left(\sigma_{z}^j
+e^{i\Omega t}|+\rangle\langle-|-e^{-i\Omega
t}|-\rangle\langle+|\right)+ \text{H.c}.
\end{eqnarray}
In the case of $\Omega\gg\{\delta, g\}$, we can omitting the fast
oscillation terms (of frequencies $\Omega\pm\delta$) and only keep
the oscillation frequency of $\delta$, then the Hamiltonian reads
 \begin{eqnarray}
 H_{2}
 =- {1 \over 2} \left(a e^{-i\delta t}+a^\dag e^{i\delta t}\right) \sum_{j=1}^N g_j
 \sigma^{z}_j.
\end{eqnarray}
Rotate back to the eigen basis of $\sigma_z$,  the Hamiltonian reads
 \begin{eqnarray}
 H_{3} \label{h3}
 =- {1 \over 2} \left(a e^{-i\delta t}+a^\dag e^{i\delta t}\right) \sum_{j=1}^N g_j
 \sigma^{x}_j.
\end{eqnarray}

For $N=2$, the corresponding closed Lie algebra for Hamiltonian
(\ref{h3}) is $\{1, a\sigma^x_j, a^\dag\sigma^x_j,
\sigma^x_1\sigma^x_2\}$. The time evolution of Hamiltonian
(\ref{h3}) is the product of their exponentials. Clearly, the first
term represents a global phase factor, and thus can be neglected.
The middle terms involve real excitation of the cavity field state,
and thus entangle the qubits with the cavity field. The last term
denotes a two qubits operation without entanglement with the cavity
field. The time evolution operator can be written as \cite{sm2}
\begin{eqnarray}
\label{u} U_2=\exp\left(-2 i A_2 \sigma_1^x\sigma_2^x\right)
\left[\prod_{j=1}^2\exp\left(-iB_{2}^j a\sigma_j^x\right)\right]
\left[\prod_{j=1}^2\exp\left(-i(B_{2}^j)^*
a^\dag\sigma_j^x\right)\right]
\end{eqnarray}
with
\begin{eqnarray} \label{parameter}
B_{2}^j={g_j \over i\delta}\left(e^{-i\delta t}-1\right),\quad
A_2={g_1g_2 \over \delta}\left[{1 \over i\delta }\left(e^{i\delta
t}-1\right)-t\right].
\end{eqnarray}
It is clear that the whole time evolution operator is not a
periodical function but $B_2^j$ is and it vanishes at intervals
\begin{equation}\label{parameter}
\delta T_n=2n\pi
\end{equation}
with $n$ being a positive integer.
At these time intervals, the time evolution operator reduce to
\begin{eqnarray}
\label{ureduce} U_2(T_n)=\exp\left[-2i A_2(T_n)
\sigma_1^x\sigma_2^x\right]
\end{eqnarray}
with $A_2(T_n)=-g_1g_2 T_n/\delta $.

The reduced operator is equivalent to a two qubits system with
Hamiltonian of the type of $\sim \sigma_1^x\sigma_2^x$. This two
qubits operation is achieved without the entanglement with the
cavity field state, and thus the cavity field do not transfer
population to the qubits system. Therefore, the operation is
insensitive to the cavity field state, the equilibrium state of
which is usually a mixed state at finite temperature. The operation
also remove the constrain of large detuning ($\delta\gg g$), and
$T_1\sim 1/g$ for $\delta\sim g$, which is comparable to the
resonant coupling strategy. This shows fast gate operation can be
achieved. Geometrically, the cavity state traverses cyclically and
returns to its original phase space coordinates at intervals $T_n$
leaving an geometric phase equals to the area of the trajectory
\cite{sm2}. This is shown to be a unconventional geometric phase
factor ~\cite{zhuunconventional},  which  still depends only on
global geometric features and is robust against random operation
errors \cite{zhuerror}, and thus high-fidelity  two-qubit operation
can be achieved \cite{f}.

This gate operation can be readily scale up to multiqubits scenario.
For the purpose of simplicity, we assume $g_j=g$ and define the
collective spin operators as $J_\nu=\Sigma_{j=1}^N \sigma_j^\nu$
with $\nu=x, y, z$. In this case, the time evolution operator can be
written as
\begin{eqnarray}
\label{un} U_N=\exp\left(-i A_N J_x^2\right)  \exp\left(-iB_N
aJ_x\right) \exp\left(-iB^*_N a^\dag J_x\right)
\end{eqnarray}
with
\begin{eqnarray} \label{parameter2}
B_N={g \over i\delta}\left(e^{-i\delta t}-1\right),\quad
A_N={g^2
\over \delta}\left[{1 \over i\delta }\left(e^{i\delta
t}-1\right)-t\right].
\end{eqnarray}
Similarly, cavity field state insensitive operator
\begin{eqnarray}
\label{unreduce} U_N(T_n)=\exp\left[-i A_N(T_n) J_x^2\right]
\end{eqnarray}
can be obtained at time intervals $\delta T_n=2n\pi$. It is obvious
that $A_N\sim A_2$, i.e., the time needed for this gate operation is
comparable to that of the two-qubit case. This is another distinct
merit of our proposed gate operation: the gate speed is not slowed
down with the increasing involved qubits. Therefore, this merit
enables efficient construction of entanglement and error correction
code \cite{zhu}.

To sum up, scheme for coupling superconducting charge qubits via a
cavity is proposed. The qubits are working at their optimal points,
the time evolution of the interaction is shown to be insensitive to
the initial state of the cavity field. This scheme enables fast gate
operation and is readily scalable to multiqubit scenario.

\bigskip

This work was supported by the NSFC (No. 11004065), the NSF of
Guangdong Province (No. 10451063101006312), and the Startup
Foundation of SCNU (No. S53005).

%\end{CJK*}

\begin{thebibliography}{99}

\bibitem{youpt} J. Q. You and F. Nori, Phys. Today  {\bf 58} (11),
42 (2005).


\bibitem{yale} A. Wallraff \emph{et al}.,
%D. I. Schuster, A. Blais, L. Frunzio, R.-S. Huang, J. Majer, S.
%Kumar, S. M. Girvin, and R. J. Schoelkopf,
Nature (London) \textbf{431}, 162 (2004); A. Blais, R.-S. Huang, A.
Wallraff, S. M. Girvin, and R. J. Schoelkopf, Phys. Rev. A {\bf 69},
062320 (2004).


\bibitem{yaledetune} J. Majer \emph{et al}., Nature (London) \textbf{449}, 443
(2007).


\bibitem{zhengdetune}  S.-B. Zheng and G.-C. Guo, Phys. Rev. Lett. \textbf{85}, 2392 (2000).

\bibitem{sm1} K. M{\o}lmer and A. S{\o}rensen, Phys. Rev. Lett.
{\bf 82}, 1835 (1999);  A. S{\o}rensen and K. M{\o}lmer,
\emph{ibid}. {\bf 82}, 1971 (1999).

\bibitem{sm2}A. S{\o}rensen and K. M{\o}lmer,  Phys. Rev. A {\bf 62}, 022311 (2000).


\bibitem{f} D. Leibfried \emph{et al}.,
%B. DeMarco, V. Meyer, D. Lucas, M. Barrett, J. Britton, W. M. Itano, B. Jelenkovic,
%C. Langer, T. Rosenband, and D. J. Wineland,
Nature (London) \textbf{422}, 412 (2003).

\bibitem{zhengdecoup} S.-B. Zheng, Phys. Rev. A  \textbf{69}, 055801 (2004).

\bibitem{zhengcavity} S.-B. Zheng, Phys. Rev. A \textbf{66}, 060303(R) (2002).



\bibitem{zhu} S.-L. Zhu, Z. D. Wang, and P. Zanardi, Phys. Rev. Lett. {\bf 94},
100502 (2005); Z.-Y. Xue and Z. D. Wang, Phys. Rev. A \textbf{75},
064303 (2007).

\bibitem{wangyd} Y.-D. Wang, A. Kemp, and K. Semba, Phys. Rev. B \textbf{79}, 024502  (2009).



\bibitem{s} E. Solano, R. L. de Matos Filho, and N. Zagury, Phys. Rev. Lett.
\textbf{90}, 027903 (2003) .

\bibitem{scully}  M. O. Scully,  and M. S. Zubairy,
\emph{Quantum optics} (Cambridge University Press, New York, 1997),
chap. 2.


\bibitem{zhuunconventional} S.-L. Zhu and Z. D. Wang, Phy. Rev. Lett. \textbf{89}, 097902 (2002).


\bibitem{zhuerror} S.-L. Zhu and P. Zanardi, Phys. Rev. A \textbf{72},
020301(R) (2005).

\end{thebibliography}
\end{document}